# In-surface confinement of topological insulator nanowire surface states

Fan W. Chen[1,2], Luis A. Jauregui[3,4], Yaohua Tan[2,3],
Michael Manfra[1,3,4,5], Gerhard Klimeck[2,3,4], Yong P. Chen[1,3,4], Tillmann Kubis[2]
[1]*Department of Physics and Astronomy, Purdue, West Lafayette, IN 47907, USA*
[2] *Network for Computational Nanotechnology, Purdue, West Lafayette, IN 47907, USA*
[3] *School of Electrical and Computer Engineering, Purdue University, West Lafayette, Indiana 47907, USA*
[4] *Birck Nanotechnology Center, Purdue University, West Lafayette, Indiana 47907, USA*
[5] *School of Materials Engineering, Purdue University, West Lafayette, Indiana 47907, USA*
*Email: fanchen@purdue.edu*

The bandstructures of [110] and [001] $Bi_2Te_3$ nanowires are solved with the atomistic 20 band tight binding functionality of NEMO5. The theoretical results reveal: The popular assumption that all topological insulator wire surfaces are equivalent is inappropriate. The Fermi velocity of chemically distinct wire surfaces differs significantly which creates an effective in-surface confinement potential. As a result, topological insulator surface states prefer specific surfaces. Therefore, experiments have to be designed carefully not to probe surfaces unfavorable to the surface states (low density of states) and thereby be insensitive to the TI-effects.

**Keywords:** Fermi Velocity, Topological Insulator, Tight binding, Band structure calculations

## I. INTRODUCTION

Topological insulator (TI) materials such as $Bi_2Te_3$ have extraordinary surface properties [1-3]. These make them a unique class of materials for applications such as low power electronic devices [4], spintronics [3], and quantum computation [5-6]. TIs host surface states with the spin perpendicular to the surface normal, spin-locked relative to the electronic in-plane momentum. Backscattering of such surface electrons requires spin-flip processes. In TI devices that are free of magnetic impurities, surface electron backscattering is therefore unlikely. Then, the surface conductance is expected to be limited by the Fermi velocity [7]. Experimental values of the Fermi velocity of $Bi_2Te_3$ surfaces show more than 25% variation [8-10]. Experiments that determine Fermi velocities and other TI surface properties are often implicitly assuming different TI surfaces host the same physics [6, 11-12]. Even many theoretical studies of TI wires assume all wire surfaces are equivalent due to rotational wire symmetry [13-15]. This assumption is only true for wires grown along [001] direction. In contrast, fabricated $Bi_2Te_3$ nanowires are grown in [110] direction and often have rectangular cross sections [16-17]. The crystal structure of [110] $Bi_2Te_3$ nanowires shows different chemical surface composition: Some surfaces are composed of Te atoms only and other surfaces contain both Te and Bi atoms. To capture this important fact of the surface chemistry requires atomistic simulations. Only then, the important effect of in-surface confinement of surface states can be simulated. It is shown in this work for $Bi_2Te_3$ nanowires that this effect confines surface states to wire surfaces with specific chemistry. It is expected that similar situations hold for other TI-materials and geometries. If experiments are set to surfaces that are unfavorable to the surface states, where the topological insulator surface states have a low density of states, the experimental setup can be effectively insensitive to the TI physics.

In this work, atomistic $sp^3d^5s^*$ (20 band, spin-orbit coupling included) tight binding bandstructure calculations of $Bi_2Te_3$ nanowires are presented. In agreement with literature the band gap of the $Bi_2Te_3$ nanowires is observed to close when the magnetic flux through the wire cross section is a half-integer flux quanta. [13-15] Deviations from literature are found in the details of the surface state energies and surface Fermi velocities: Fermi velocities of chemically different surfaces differ and create an effective surface-state confining potential around the wire surface. Guided by the atomistic results, the analytical Fermi velocity model of Ref. [13] is augmented to cover these differences of the wire surface chemistry.

In section II, the two methods used in this work are presented. This covers the atomistic tight binding features of NEMO5 and the analytical model of Ref. [13] augmented to cover variations in the wire surface chemistry. In section III, the atomistic tight binding bandstructure results of NEMO5 for $Bi_2Te_3$ nanowires in the presence of magnetic fields are verified against literature [13]. Bandstructures of rectangular $Bi_2Te_3$ nanowires with different ratios of pure Te and mixed atom type surfaces are presented then. These bandstructures serve as fitting targets for the surface Fermi velocities of the analytical model. Confinement effects of the surface states are shown after that. The analytical model is then used to explain this confinement of the wire surface states. Finally the paper concludes with a summary of the finding.

## II. METHOD

In this work, atomistic sp3d5s* (20 band, spin-orbit coupling included) tight binding bandstructure calculations of $Bi_2Te_3$ nanowires are calculated with the multipurpose NanoElectronics Modelling Tool (NEMO5) [18-19]. A quintuple layer of $Bi_2Te_3$ consists of a sequence of five atomic layers: Te1, Bi, Te2, Bi, and Te1. "Te1" and "Te2" both denote Tellurium, but they differ in the chemical surrounding: The neighbor layers of Te1 consist of Te1 and Bi, whereas the Te2 atom layer lies between two Bi atom layers. Tight binding parameters for $Bi_2Te_3$ are taken from Ref. [20]. Pairs of degenerate states are combined into symmetric and anti-symmetric states. Magnetic fields are included with the Peierl's phase in symmetric gauge [21]. All presented atomistic calculations are numerically very intense and require typically about one million CPUs on the Blue Waters supercomputer.

To ease understanding of the tight binding results, the analytical model of Refs. [13-15] is augmented to support nanowires that are not rotationally symmetric along the transport axis. Typical examples for such wires are grown along [110] direction and are rectangular in the cross section. Such wires have two different types of facets: one type contains atoms of all types ("mixed surface"), while the second consists of Te1 atoms (referred as "Te1 surface"). For these surface types, different Fermi velocities are assumed: $v_{f1}$ for the mixed surface and $v_{f2}$ for the pure Te1 one. The energy difference of the wire surface states ΔE with vanishing momentum ($k = 0$), is assumed to be

$$\Delta E = \begin{cases} \frac{v_{f1}h}{P} \quad (Type\ I, W > T) \\ \frac{v_{f2}h}{P} \quad (Type\ II, T > W) \end{cases} \qquad (1)$$

Here, $P$ is the perimeter of the wire, equals to (2W+2T); W is the dimension of one mixed surface ("width of the wire") and T is the dimension of one pure Te1 surface ("thickness of the wire"). The two Fermi velocities are fit to match the surface state quantization of the tight binding results.

## III. RESULTS

If not stated otherwise: all figures show tight binding results. All considered $Bi_2Te_3$ wires are grown along [110] direction.

It is an accepted rule in literature that the band gap of TI nanowires closes when the magnetic flux through the wire cross section agrees with half integer multiples of the magnetic flux quantum ($\Phi_0 = h/e$). The largest band gap of the TI wires is expected with magnetic fluxes equal to integer multiples of the flux quantum. However, this knowledge is based on non-atomistic models (i.e. envelope function approximations). Figures 1 show the atomistic tight binding bandstructures of $12 \times 48 nm^2$ $Bi_2Te_3$ nanowires with varying magnetic fields along the wire growth direction. Here, the smaller facets are pure Te1 type. The atomistic calculations indeed follow the rule of vanishing and maximal band gaps as a function of the magnetic flux. Equivalent behavior was observed for atomistic tight binding calculations of $Bi_2Te_3$ nanowires for a great variety of cross sections (ranging from $6 \times 24 nm^2$ to $60 \times 150 nm^2$). Atomistic tight binding calculations showed that different geometries and facet configurations do not alter the rule for band gap maxima and minima.

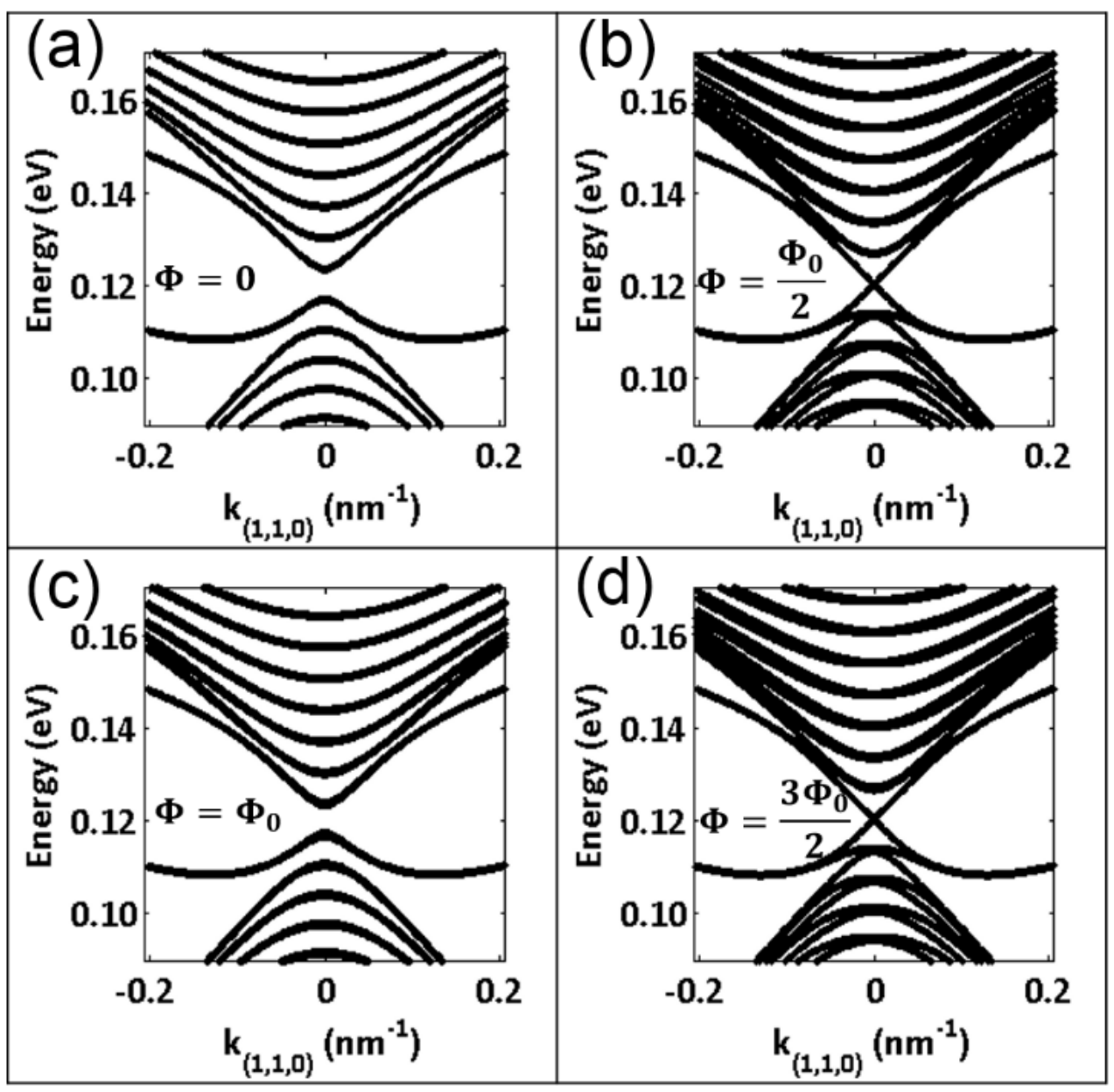

FIG. 1. Bandstructures of 12x48 $nm^2$ $Bi_2Te_3$ nanowires (type I) with varying magnetic field along the wire axis. The wire banstructure without a magnetic field (a) or with a magnetic field corresponding to the magnetic flux quantum (c) has the largest band gap and every state is double degenerate. Bandstructures with magnetic fields corresponding to 0.5 (b) and 1.5 (d) magnetic flux quanta have disappearing band gap and are only degenerate at k=0.

Although different wire configurations follow the same rule for the band gap with magnetic fields, the band structure details depend significantly on the ratio of pure Te1 and mixed facet dimensions. This is exemplified in Figs. 2: Figures 2a) compare the atomistic structure of two $12 \times 48 nm^2$ $Bi_2Te_3$ nanowires that differ in the size of the pure Te1 and mixed facets. For later reference, wires with larger mixed than pure Te1 facets are termed "Type I", the other cases as "Type II". The bandstructures of the two cases in Figs. 2b) show an energy difference of the wire surface states with vanishing momentum ΔE of 6.2meV for the type I and 10.5meV for the type II nanowire of Figs. 2a). This difference in ΔE is in contrast to non-atomistic models that cannot distinguish wires of type I and type II.

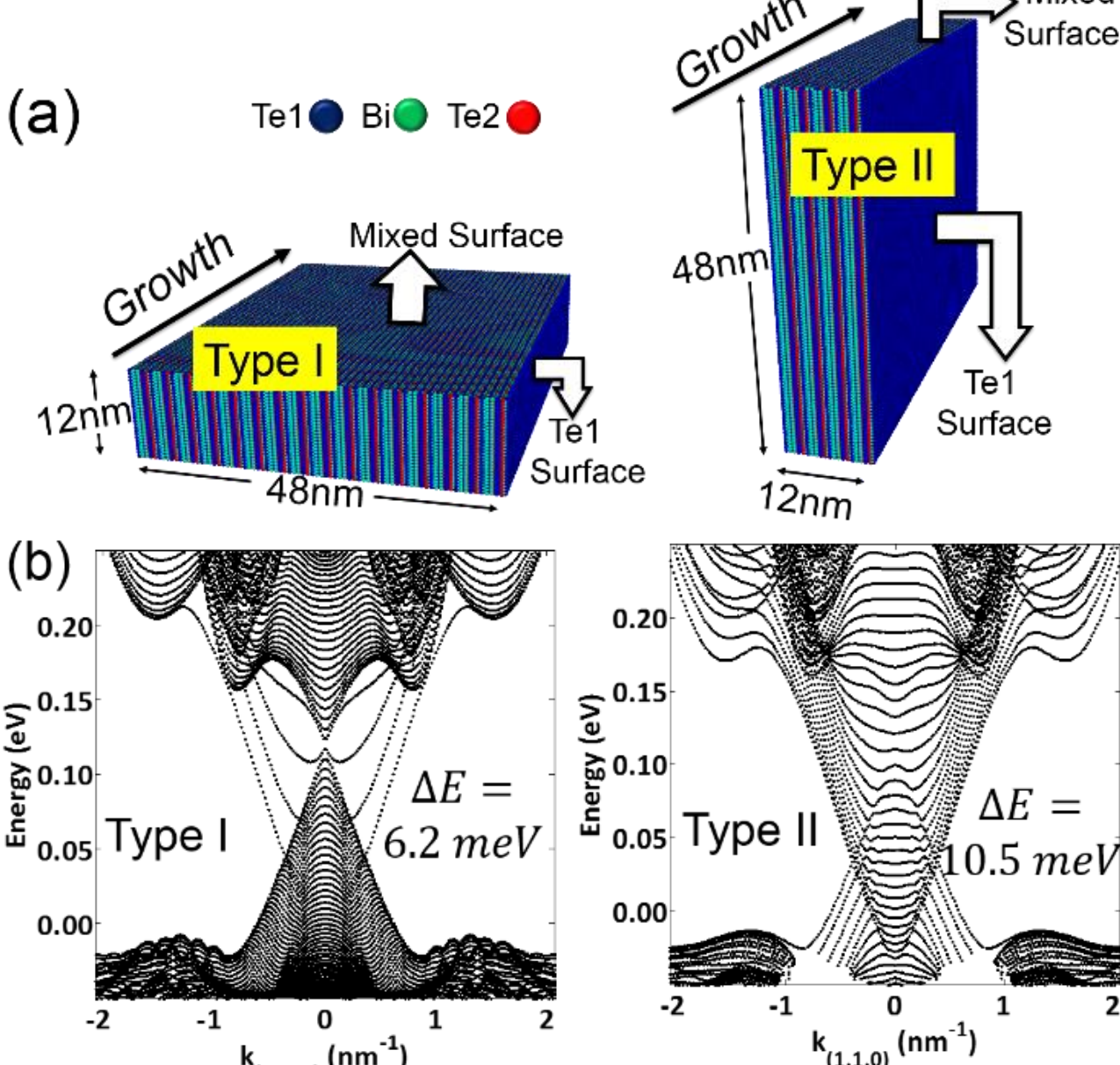

FIG. 2. (a) Atomic structures of 12x48 nm$^2$ $Bi_2Te_3$ nanowires grown along the [110] direction in the two possible surface configurations. In the type I wire, the pure Te1 facet is much smaller than in the type II configuration. (b) Bandstructures of the nanowires of (a) show different quantization energies ΔE for the surface states at k=0.

Figures 3 show the ΔE as a function of the perimeter of $Bi_2Te_3$ nanowires grown in [110] (a) and [001] (b) direction. In all cases, the longer edge of the wire cross section is kept constant to 48nm, while the smaller edge varies. For [110] grown wires of type I, the mixed surface is kept constant, while for type II wires, the pure Te1 surface is constant. Please note that all surfaces of [001] wires are of mixed atom type and equivalent. Therefore, a type I and type II distinction is meaningless for [001] wires, i.e. the surfaces A and B are equivalent. Therefore, nanowires grown in [110] direction show a strong dependence of ΔE on the facets ratio configuration, while ΔE of [001] wires is virtually independent of that. Due to the linear nature of all results shown in Figs.3, the analytical model discussed in the method section gives an almost perfect fit to the numerical atomistic tight binding data. It turns out, the Fermi velocities for the [110] wires are $v_{f1} = 1.28 \times 10^5 m/s$ and $v_{f2} = 4.36 \times 10^5 m/s$ and for the [001] wires $v_{f1} = v_{f2} = 1.28 \times 10^5 m/s$. Note that mixed surfaces that are chemically equivalent (i.e. having the same portion of Te1, Te2 and Bi atoms) have always the same Fermi velocity, irrespective of the wire growth direction.

The values of the Fermi velocities $v_{f1}$ and $v_{f2}$ vary with the size of the bigger facet (kept constant in Figs. 3). For instance, if the bigger facet has the dimension of 120 nm the following Fermi velocities are found: $v_{f1} = 1.39 \times 10^5 m/s$ and $v_{f2} = 4.50 \times 10^5 m/s$.

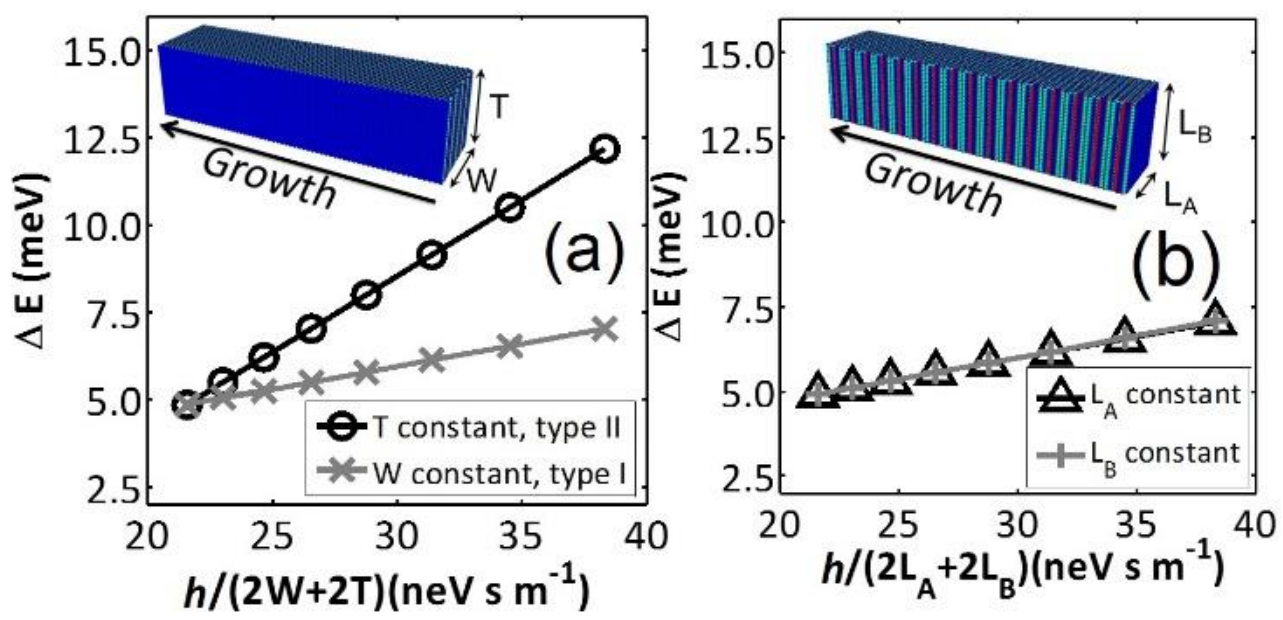

FIG. 3. Surface state quantization energies ΔE as a function of the inverse wire perimeter for $Bi_2Te_3$ nanowires grown in [110] direction (a) and in [001] direction (b). Different facets of [110] wires differ in their chemistry and give different quantization energies. This is in contrast to [001] wires. Surface Fermi velocities result from linear approximations to these data as discussed in the main text.

This change in the Fermi velocity indicates confinement effects within the wire surface. This in-surface confinement is illustrated in Figs.4 which show the absolute squared wavefunctions for the first 3 surface states of the type I nanowire in Figs. 2 with energies above about 0.12eV and momentum k=0.025nm$^{-1}$. The surface states are mainly located at the mixed type facets. The number of minima of the surface states envelope (shown in Fig. 4b) increases with the state's energy – similar to confined electronic states in quantum wells. The calculations also show stronger in-surface confinement effect with increasing momentum. The in-surface confinement vanishes at the Γ-point. This finding can be understood with the analytical model of Eq.(1): The dispersion difference of the two different nanowire facets (pure Te1 and mixed type) yields an effective, momentum dependent potential between the facet types (see schematic of Fig.5a). This potential vanishes at the Γ-point and increases with finite momenta. This potential effectively creates a system of 2 quantum wells within the wire surface (see schematic of Fig.5b). The surface states envelopes' confinement is typical for such quantum wells.

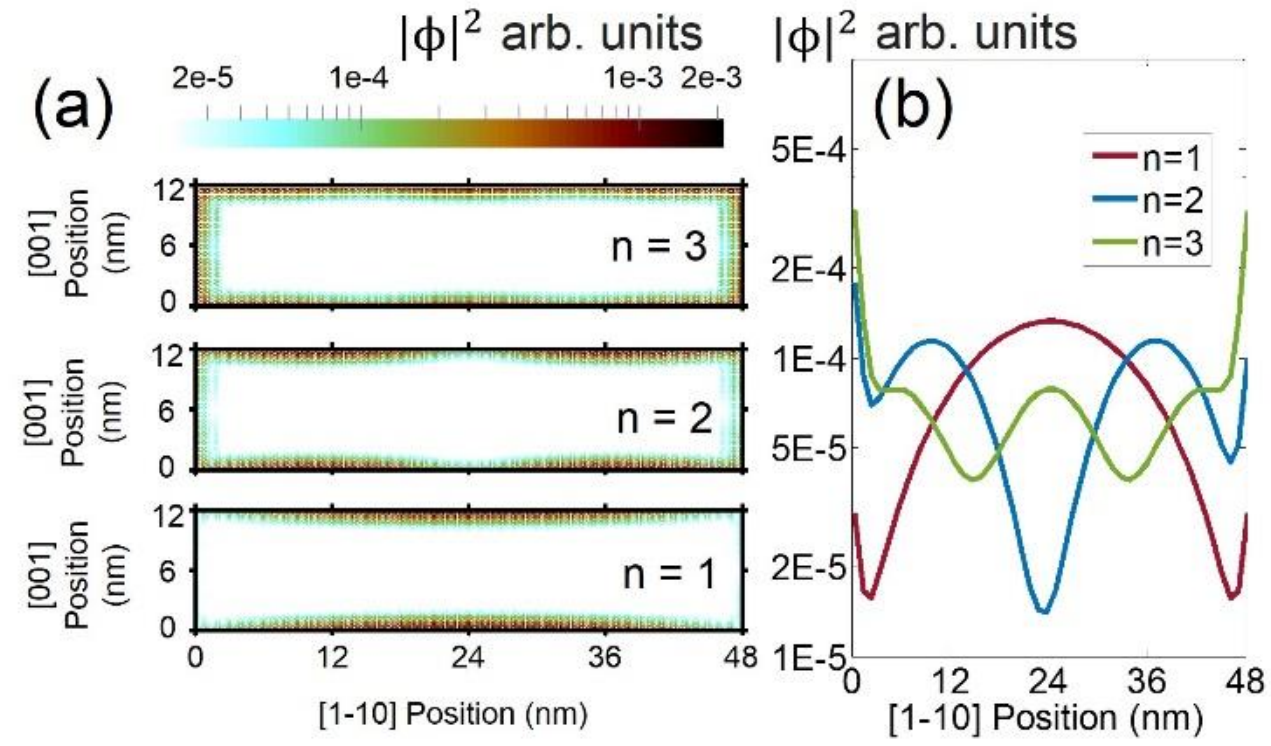

FIG. 4. (a) Contour plot of the absolute squared wavefunctions of the 3 surface states right above the Dirac point of the type I $Bi_2Te_3$ nanowire of Fig.2(a) for the momentum k=0.025 nm$^{-1}$. (b) Unit-cell average of the surface states in (a) along the [001] wire coordinate. "n" represents the subband index and the number of maxima of the surface states envelope.

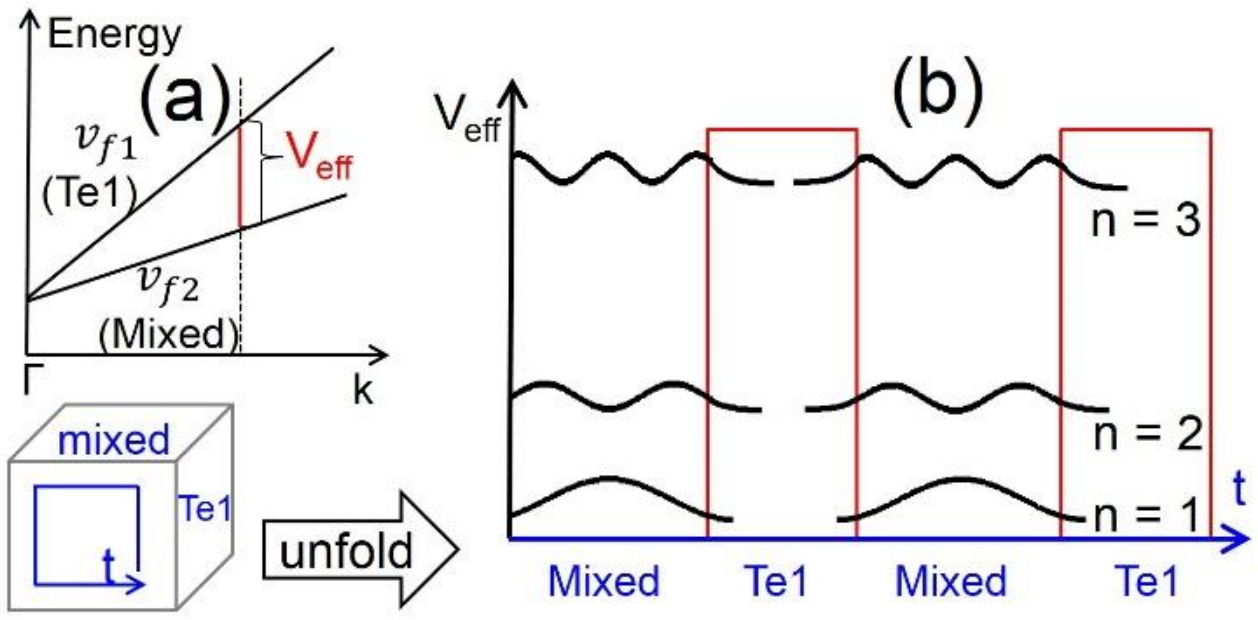

FIG. 5. (a) Schematic surface state bandstructure of $Bi_2Te_3$ nanowires grown in (110) direction as shown in Fig.2(b) with different Fermi velocities for the pure Te1 and the mixed surfaces. The different Fermi velocities cause an effective potential offset between the two surfaces for finite momenta (highlighted with $V_{eff}$). (b) The effective potential $V_{eff}$ (red line) of (a) along the unfolded nanowire perimeter confines the surface states predominantly on the mixed surface (illustrated with schematic wavefunctions in black).

## IV. CONCLUSION

In this work, NEMO5's atomistic tight binding models are applied on $Bi_2Te_3$ nanowire bandstructures for wires grown along [110] and [001] direction. The atomistic representation unveils for experimentally common, rectangular [110] nanowires two chemically different types of surfaces. These surfaces host topological insulator surface states with different Fermi velocities. The surface states spread over all facets, but they are subject to momentum-dependent in-surface confinement. This finding is critical for experiments on TI-surfaces: TI-properties should be measured only on TI-favorable surfaces. It is imaginable that variations of measured Fermi velocities in $Bi_2Te_3$ might trace back to the different Fermi velocities of the different surface kinds. This situation is different in [001] wires due to their chemically equivalent facets. Both situations can be well reproduced with an analytical model presented in this work as well.

## V. ACKNOWLEDGEMENTS

We acknowledge discussion with Sicong Chen and James Charles as well as support from Blue Waters sustained-petascale computing project (awards OCI-0725070 and ACI-1238993) and the state of Illinois. This work was also supported by the Semiconductor Research Corporation's (SRC task 2141) Nanoelectronics Research Initiative and National Institute of Standards & Technology through the Midwest Institute for Nanoelectronics Discovery (MIND), and the Intel Corporation.

[1] X. Qi, and S. Zhang, *Reviews of Modern Physics* **83**, 1057 (2011).
[2] M. Z. Hasan, and C. L. Kane, *Reviews of Modern Physics* **82**, 3045 (2010).
[3] J. E. Moore, *Nature* **464**, 194 (2010).
[4] C. Z. Chang *et al.*, *Science* **340**, 167 (2013).
[5] A. Cook, and M. Franz, *Phys. Rev. B* **84**, 201105 (2011).
[6] L. Fu, and C. L. Kane, *Phys. Rev. Lett.* **100**, 96407 (2008).
[7] C. Hwang, D. A. Siegel, S. Mo, W. Regan, A. Ismach, Y. Zhang, A. Zettl, and A. Lanzara, *Scientific Reports* **2**, 590 (2012).
[8] K. Kuroda, *et al.*, *Phys. Rev. Lett.* **105**, 076802 (2010).
[9] D. Qu, Y. S. Hor, J. Xiong, R. J. Cava, and N. P. Ong, *Science* **329**, 821 (2010).
[10] T. Zhang, *et al.*, *Phys. Rev. Lett.* **103**, 266803 (2009).
[11] D. Hsieh, D. Qian, L. Wray, Y. Xia, Y. S. Hor, R. J. Cava, and M. Z. Hasan, *Nature* **452**, 970 (2008).
[12] H. Peng, K. Lai, D. Kong, S. Meister, Y. Chen, X. Qi, S. Zhang, Z. Shen, and Y. Cui, *Nature Materials* ***9***, 225 (2010).
[13] J. H. Bardarson, P. W. Brouwer, and J. E. Moore, *Phys. Rev. Lett.* **105**, 156803 (2010).
[14] G. Rosenberg, H-M. Guo, and M. Franz, *Phys. Rev. B* **82**, 041104 (2010).
[15] Y. Zhang, and A. Vishwanath, *Phys. Rev. Lett.* **105**, 206601 (2010).
[16] F. Xiu *et al., Nature Nanotechnology* **6**, 216 (2011).
[17] L. A. Jauregui, M. T. Pettes, L. P. Rokhinson, L. Shi, and Y. P. Chen, *Scientific Reports* **5**, 8452 (2015).
[18] S. Steiger, M. Povolotskyi, H. H. Park, T. Kubis, and G. Klimeck, *IEEE Trans. Nanotechnol*. **10**, 1464 (2011).
[19] R. Lake, G. Klimeck, R. C. Bowen, and D. Jovanovic, *J. Appl. Phys.* **81**, 7845 (1997).
[20] S. Lee, and P. Allmen, *App. Phys. Lett.* **88**, 022107 (2006).
[21] M. Graf, and P. Vogl, *Phys. Rev. B* **51**, 4940 (1995).